\documentclass[pre,epsfig,floats,twocolumn,amssymb,amsmath,floatfix,showpacs]{revtex4}
\usepackage{graphicx}

\begin{document}

\title{Analytic results for the three-sphere swimmer at low Reynolds number}

\author{Ramin Golestanian}
\email{r.golestanian@sheffield.ac.uk} \affiliation{Department of
Physics and Astronomy, University of Sheffield, Sheffield S3 7RH,
UK}

\author{Armand Ajdari}
\affiliation{Gulliver, UMR CNRS 7083, ESPCI, 10 rue Vauquelin, 75005
Paris, France}

\date{\today}

\begin{abstract}
The simple model of a low Reynolds number swimmer made from three
spheres that are connected by two arms is considered in its general
form and analyzed. The swimming velocity, force--velocity response,
power consumption, and efficiency of the swimmer are calculated both
for general deformations and also for specific model prescriptions.
The role of noise and coherence in the stroke cycle is also
discussed.
\end{abstract}

\maketitle

\section{Introduction}  \label{sec:intro}

There is a significant complication in designing swimmers at small
scale as they have to undergo non-reciprocal deformations to break
the time-reversal symmetry and achieve propulsion at low Reynolds
number \cite{taylor}. While it is not so difficult to imagine
constructing motion cycles with the desired property when we have a
large number of degrees of freedom at hand---like nature does, for
example---this will prove nontrivial when we want to design
something with only a few degrees of freedom and strike a balance
between simplicity and functionality, like most human-engineered
devices \cite{purcell1}. Recently, there has been an increased
interest in such designs
\cite{3SS,avron,dreyfus0,kulic,lee,NG,feld,gauger,lesh,tam,earl,gareth,pooley,lauga,Kruse}
and an interesting example of such robotic micro-swimmers has been
realized experimentally using magnetic colloids attached by
DNA-linkers \cite{Dreyfus}. While constructing small swimmers that
generate surface distortions is a natural choice, it is also
possible to take advantage of the general class of phoretic
phenomena to achieve locomotion, as they become predominant at small
scales \cite{phoretic}.

Here we consider a recently introduced model for a simple low
Reynolds number swimmer that is made of three linked spheres
\cite{3SS}, and present a detailed analysis of it motion. Unlike the
Purcell swimmer \cite{purcell1} that is difficult to analyze because
it takes advantage of the rotational degrees of freedom of finite
rods that move near each other \cite{howard1}, the three-sphere
swimmer model is amenable to analytical analysis as it involves the
translational degrees of freedom in one dimension only, which
simplifies the tensorial structure of the fluid motion. We present
closed form expressions for the swimming velocity with arbitrary
swimming deformation cycles, and also use a perturbation scheme to
simplify the results so that the study can be taken further. We
examine various mechanical aspects of the motion including the
pattern of the internal forces during the swimming, the
force--velocity response of the swimmer due to external loads, the
power consumption rate, and the hydrodynamic efficiency of the
swimmer. Finally, we consider the role of the phase difference
between the motion of the two parts of the swimmer and propose a
mechanism to build in a constant (coherent) phase difference in a
system that is triggered from the outside. We also discuss the
effect of noise on the swimming velocity of the model system. We
also note that the three-sphere low Reynolds swimmer has been
recently generalized to the case of a swimmer with a macroscopic
cargo container \cite{cargo}, and a swimmer whose deformations are
driven by stochastic random configurational transitions
\cite{chem3SS}.

The rest of the paper is organized as follows. In Sec.
\ref{sec:simpl} the three-sphere low Reynolds number swimmer model
is introduced in a general form and a simplified analysis of its
swimming is presented. This is followed by a detailed discussion of
its swimming velocity in Sec. \ref{sec:velocity}, with a focus on a
few particular examples of swimming stroke cycles. Section
\ref{sec:int-force} is devoted to a discussion on internal stresses
and forces acting during the swimming cycle, and Sec.
\ref{sec:force-vel} studies the force--velocity response of the
swimmer. The power consumption and efficiency of the model swimmer
are discussed in Sec. \ref{sec:power}, followed by concluding
remarks in Sec. \ref{sec:conc}. Appendix \ref{app:V-full} contains
the closed form expression for the swimming velocity of the general
asymmetric swimmer, which is the basis of some of the results
discussed in the paper.

\section{Three-Sphere Swimmer: Simplified Analysis}
\label{sec:simpl}

\begin{figure}[b]
\includegraphics[width=.9\columnwidth]{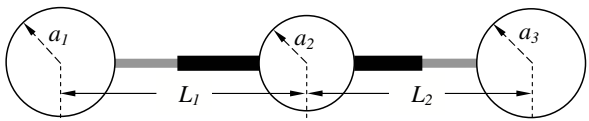}
\caption{Schematics of the three-sphere swimmer. The two arms can
open and close in a prescribed form, and this could lead to
locomotion if the swimming strokes are nonreciprocal. }
\label{fig:3SS-schem}
\end{figure}

We begin with a simplified model geometry which consists of three
spheres of radii $a_i$ ($i=1,2,3$) that are separated by two arms of
lengths $L_1$ and $L_2$ as depicted in Fig. \ref{fig:3SS-schem}.
Each sphere exerts a force $f_i$ on (and experiences a force $-f_i$
from) the fluid that we assume to be along the swimmer axis. In the
limit $a_i/L_j \ll 1$, we can use the Oseen tensor \cite{lrnh,Oseen}
to relate the forces and the velocities as

\begin{eqnarray}
v_1&=&\frac{f_1}{6 \pi \eta a_1}+\frac{f_2}{4 \pi \eta L_1}
+\frac{f_3}{4 \pi \eta (L_1+L_2)}, \label{v1-fi} \\
v_2&=&\frac{f_1}{4 \pi \eta L_1}+\frac{f_2}{6 \pi \eta a_2}
+\frac{f_3}{4 \pi \eta L_2}, \label{v2-fi} \\
v_3&=&\frac{f_1}{4 \pi \eta (L_1+L_2)}+\frac{f_2}{4 \pi \eta
L_2}+\frac{f_3}{6 \pi \eta a_3}. \label{v3-fi}
\end{eqnarray}
Note that in this simple one dimensional case, the tensorial
structure of the hydrodynamic Green's function (Oseen tensor) does
not enter the calculations as all the forces and velocities are
parallel to each other and to the position vectors. The swimming
velocity of the whole object is the mean translational velocity,
namely
\begin{equation}
V=\frac{1}{3} (v_1+v_2+v_3).\label{V-def}
\end{equation}
We are seeking to study autonomous net swimming, which requires the
whole system to be force-free (i.e. there are no external forces
acting on the spheres). This means that the above equations are
subject to the constraint
\begin{equation}
f_1+f_2+f_3=0.\label{force-free-def}
\end{equation}
Eliminating $f_2$ using Eq. (\ref{force-free-def}), we can
calculate the swimming velocity from Eqs. (\ref{v1-fi}),
(\ref{v2-fi}), (\ref{v3-fi}), and (\ref{V-def}) as
\begin{eqnarray}
V_0&=&\frac{1}{3} \left[\left(\frac{1}{a_1}-\frac{1}{a_2}\right)
+\frac{3}{2} \left(\frac{1}{L_1+L_2}-\frac{1}{L_2}\right)\right]
\left(\frac{f_1}{6 \pi
\eta}\right)\nonumber \\
&+&\frac{1}{3} \left[\left(\frac{1}{a_3}-\frac{1}{a_2}\right)
+\frac{3}{2} \left(\frac{1}{L_1+L_2}-\frac{1}{L_1}\right)\right]
\left(\frac{f_3}{6 \pi \eta}\right), \nonumber \\
\label{V-f1-f3}
\end{eqnarray}
where the subscript $0$ denotes the force-free condition. To close
the system of equations, we should either prescribe the forces
(stresses) acting across each linker, or alternatively the opening
and closing motion of each arm as a function of time. We choose to
prescribe the motion of the arms connecting the three spheres, and
assume that the velocities
\begin{eqnarray}
{\dot L}_1&=& v_2-v_1,\label{L-dot-1-def} \\
{\dot L}_2&=& v_3-v_2,\label{L-dot-2-def}
\end{eqnarray}
are known functions. We then use Eqs. (\ref{v1-fi}), (\ref{v2-fi}),
(\ref{v3-fi}), and (\ref{force-free-def}) to solve for $f_1$ and
$f_3$ as a function of ${\dot L}_1$ and ${\dot L}_2$. Putting the
resulting expressions for $f_1$ and $f_3$ back in Eq.
(\ref{V-f1-f3}), and keeping only terms in the leading order in
$a_i/L_j$ consistent with our original scheme, we find the average
swimming velocity to the leading order.

\section{Swimming Velocity}
\label{sec:velocity}

The result of the above calculations is the lengthy expression of
Eq. (\ref{V-full-1}) reported in Appendix \ref{app:V-full}. This
result is suitable for numerical studies of swimming cycles with
arbitrarily large deformations. For the simple case where all the
spheres have the same radii, namely $a=a_1=a_2=a_3$, Eq.
(\ref{V-f1-f3}) simplifies to
\begin{equation}
V_0=\frac{a}{6} \left[\left(\frac{{\dot L}_2-{\dot
L}_1}{L_1+L_2}\right)+2 \left(\frac{{\dot L}_1}{L_2}-\frac{{\dot
L}_2}{L_1}\right)\right],\label{eq:v0-a-eq-large-amp}
\end{equation}
plus terms that average to zero over a full swimming cycle. Equation
\ref{eq:v0-a-eq-large-amp} is also valid for arbitrarily large
deformations.

We can also consider relatively small deformations and perform an
expansion of the swimming velocity to the leading order. Using
\begin{eqnarray}
L_1&=&\ell_1+u_1, \label{L-1-u-1} \\
L_2&=&\ell_2+u_2, \label{L-2-u-2}
\end{eqnarray}
in Eq. (\ref{V-full-1}), and expanding to the leading order in
$u_i/\ell_j$, we find the average swimming velocity as
\begin{equation}
\overline{V_0}=\frac{K}{2} \;\overline{(u_1 \dot{u}_2-\dot{u}_1
u_2)},\label{V-bar-1}
\end{equation}
where
\begin{equation}
K=\frac{3 \;a_1 a_2 a_3}{(a_1+a_2+a_3)^2}
\left[\frac{1}{\ell_1^2}+\frac{1}{\ell_2^2}-\frac{1}{(\ell_1+\ell_2)^2}\right].\label{eq:K-def-1}
\end{equation}
In the above result, the averaging is performed by time integration
in a full cycle. Note that terms proportional to $u_1 \dot{u}_1$,
$u_2 \dot{u}_2$, and $u_1 \dot{u}_2+\dot{u}_1 u_2$ are eliminated
because they are full time derivatives and they average out to zero
in a cycle. Equation (\ref{V-bar-1}) clearly shows that the average
swimming velocity is proportional to the enclosed area that is swept
in a full cycle in the configuration space [i.e. in the $(u_1,u_2)$
space]. This result, which is valid within the perturbation theory,
is inherently related to the geometrical structure of theory the low
Reynolds number swimming studied by Shapere and Wilczek
\cite{geometry}. Naturally, the swimmer can achieve higher
velocities if it can maximize this area by introducing sufficient
phase difference between the two deformation cycles (see below). We
also note that the above result is more general than what was
previously considered in Ref. \cite{3SS}, which corresponded to the
class of configurational changes that happen one at a time, i.e.
spanning rectangular areas in the configuration space \cite{note}.

We can actually obtain Eq. (\ref{V-bar-1}) from a rather general
argument. Since the deformation of the arms is prescribed, the
instantaneous net displacement velocity of the swimmer should take
on a series expansion form of
\begin{math}
v(t)=A_i {\dot u}_i+B_{ij} {\dot u}_i {u}_j+C_{ijk} {\dot u}_i {u}_j
{u}_k+\cdots,
\end{math}
where the coefficients $A_i$, $B_{ij}$, $C_{ijk}$, {\em etc.} are
purely geometrical prefactors ({\em i.e.} involving only the length
scales $a_i$'s and $\ell_i$'s). Terms of higher order than one in
velocity will have to be excluded on the grounds that in Stokes
hydrodynamics forces are linearly dependent on (prescribed)
velocities, and then the velocity any where else is also linearly
proportional to the forces, which renders an overall linear
dependency of the swimming velocity on set velocities. Moreover,
higher order terms in velocity would require a time scale such as
the period of the motion to balance the dimensions, which is not a
quantity that is known to the system at any instant (i.e. would
require nonlocal effects). Since the motion is periodic and we
should average over one complete cycle to find the net swimming
velocity, we can note that the only combination that survives the
averaging process up to the second order is $u_1 \dot{u}_2-\dot{u}_1
u_2$, which yields Eq. (\ref{V-bar-1}). Note that this argument
works even if the spheres have finite large radii that are not small
comparable to the average length of the arms.

It is instructive at this point to examine a few explicit examples
of swimming cycles for the three-sphere swimmer.

\subsection{Harmonic Deformations}\label{sec:harmonic}

Let us consider harmonic deformations for the two arms, with
identical frequencies $\omega$ and a mismatch in phases, namely
\begin{eqnarray}
u_1(t)=d_1 \cos(\omega t+\varphi_1), \label{u-1-harm} \\
u_2(t)=d_2 \cos(\omega t+\varphi_2). \label{u-2-harm}
\end{eqnarray}
The average swimming velocity from Eq. (\ref{V-bar-1}) reads
\begin{equation}
\overline{V_0}=\frac{K}{2}\;d_1 d_2 \omega
\sin(\varphi_1-\varphi_2).\label{V-bar-harm}
\end{equation}
This result shows that the maximum velocity is obtained when the
phase difference is $\pi/2$, which supports the picture of
maximizing the area covered by the trajectory of the swimming cycle
in the parameter space of the deformations. A phase difference of
$0$ or $\pi$, for example, will create closed trajectories with zero
area, or just lines.

\subsection{Simultaneous Switching and Asymmetric Relaxation}\label{sec:exponential}

Thinking about practical aspects of implementing such swimming
strokes in real systems, it might appear difficult to incorporate a
phase difference in the motion of the two parts of a swimmer. In
particular, for small scale swimmers we would not have direct
mechanical access to the different parts of the system and the
deformations would be more easily triggered externally by some kind
of generic interaction with the system, such as shining laser
pulses. In this case, we need to incorporate a net phase difference
in the response of the two parts of the system to simultaneous
triggers. This can be achieved if the two parts of the system have
different relaxation times. To illustrate this, imagine that the
arms of the swimmer could switch their lengths via exponential
relaxation between two values of $\ell_i-d_i/2$ and $\ell_i+d_i/2$
back and forth as a switch is turned on and off. The deformation can
be written as
\begin{equation}
u_i(t)=\left\{\begin{array}{ll} d_i
\left[-\frac{1}{2}+\frac{1-e^{-t/\tau_i}}
{1-e^{-T/2\tau_i}}\right], & \; 0<t<\frac{T}{2},  \\ \\
d_i
\left[-\frac{1}{2}+\frac{e^{-t/\tau_i}-e^{-T/\tau_i}}
{e^{-T/2\tau_i}-e^{-T/\tau_i}}\right],
& \; \frac{T}{2}<t<T,
\end{array} \right. \label{u-i-harm}
\end{equation}
for $i=1,2$, where $\tau_i$'s are the corresponding relaxation
times, and $T$ is the (common) period of the switchings. We find
\begin{equation}
\overline{V_0}=\frac{K}{8} \;d_1 d_2
\left(\frac{1}{\tau_1}-\frac{1}{\tau_2}\right) \;{\cal
F}\left(\frac{T}{4 \tau_1},\frac{T}{4
\tau_2}\right),\label{V-bar-expo}
\end{equation}
where
\begin{equation}
{\cal F}(x,y)=\frac{1}{\sinh x \; \sinh y}
\left[\frac{\sinh(x+y)}{(x+y)}-\frac{\sinh(x-y)}{(x-y)}\right].\label{f(x,y)}
\end{equation}
The above function is a smooth and monotonically decaying function
of both $x$ and $y$ that is always positive, and it has the
asymptotic limits ${\cal F}(0,0)=\frac{2}{3}$, ${\cal F}(x,y \to
\infty)=0$, and ${\cal F}(x \to \infty,y)=0$. Here, the phase
mismatch is materialized in the difference in the relaxation times,
despite the fact that the deformations are switched on and off
simultaneously.

\subsection{Noisy Deformations}\label{sec:noisy}

Another important issue in practical situations is the inevitability
of random or stochastic behavior of the deformations. In small
scales, Brownian agitations of the environment become a significant
issue, and we would like to know how feasible it is to extract a net
coordinated swimming motion from a set of two noisy deformation
patterns. Using the Fourier transform of the deformations ($i=1,2$)
\begin{equation}
u_i(t)=\int \frac{d \omega}{2 \pi} \; u_i(\omega) e^{i \omega t},
\label{u-i-1-noise}
\end{equation}
we can calculate the time-averaged swimming velocity of Eq.
(\ref{V-bar-1}) as
\begin{equation}
\overline{V_0}=\frac{K}{2}\; \frac{1}{T} \int \frac{d \omega}{2
\pi}\; i \omega
\overline{\left[u_2(\omega)u_1(-\omega)-u_1(\omega)u_2(-\omega)\right]}.
\label{V-bar-noise-1}
\end{equation}
For deformations that have discrete spectra ($i=1,2$), namely
\begin{equation}
u_i(t)=\sum_n \; d_{i n} \cos(\omega_n t+\varphi_{i n}),
\label{u-i-2-noise}
\end{equation}
we find
\begin{equation}
\overline{V_0}=\frac{K}{2} \sum_n \; \overline{d_{1 n} d_{2 n}
\omega_n \sin(\varphi_{1 n}-\varphi_{2 n})}.\label{V-bar-noise-2}
\end{equation}
This shows that a net swimming is the result of coordinated motions
of the different modes in the frequency spectrum, and the net
velocity is the sum of the individual contributions of the different
modes. As long as we can achieve a certain degree of coherence in a
number of selected frequencies, we can have a net swimming despite
the noisy nature of the deformations.

\section{Internal Forces and Stresses}\label{sec:int-force}

\begin{figure}
\includegraphics[width=.9\columnwidth]{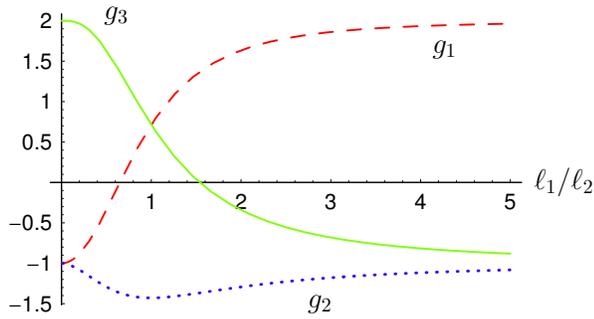}
\caption{(color online.) The dimensionless forces exerted on the
fluid at the locations of the spheres [defined in Eq.
(\ref{eq:g-def})] as functions of the relative size of the two arms.
Dashed line (red) corresponds to Sphere 1, dotted line (blue)
corresponds to Sphere 2, and solid line (green) corresponds to
Sphere 3.} \label{fig:g1g2g3}
\end{figure}

The interaction of the spheres and the medium involves forces as
these parts of the swimmer make their ways through the viscous fluid
when performing the swimming strokes. We have calculated these
forces for a general swimmer, but since their expressions are quite
lengthy, we choose to present them only in the particular case where
all the spheres have equal radii, namely $a_1=a_2=a_3=a$. For
arbitrarily large deformations, we find
\begin{eqnarray}
f_1&=&\frac{\pi \eta a^2}{2}\left[4 \;\frac{{\dot L}_1}{L_2}+2\;
\frac{{\dot L}_2}{L_1}+\frac{{\dot L}_1-{\dot
L}_2}{L_1+L_2}\right], \label{f1-arb} \\
f_2&=&\frac{\pi \eta a^2}{2}\left[-2\; \frac{{\dot L}_1}{L_2}+2\;
\frac{{\dot L}_2}{L_1}-2\;\frac{{\dot L}_1-{\dot
L}_2}{L_1+L_2}\right], \label{f2-arb} \\
f_3&=&\frac{\pi \eta a^2}{2}\left[-2\; \frac{{\dot L}_1}{L_2}-4\;
\frac{{\dot L}_2}{L_1}+\frac{{\dot L}_1-{\dot L}_2}{L_1+L_2}\right],
\label{f3-arb}
\end{eqnarray}
up to terms that average to zero over a full cycle. One can check
that the above expressions manifestly add up to zero, as they
should. For small relative deformations, we find the following
expressions for the average forces exerted on the fluid
\begin{eqnarray}
\overline{f_1}&=&\frac{\pi \eta a^2}{2}\left[-\frac{1}{\ell_1^2}
+\frac{2}{\ell_2^2}+\frac{1}{(\ell_1+\ell_2)^2}\right]
\overline{(u_1 \dot{u}_2-\dot{u}_1 u_2)}, \nonumber \\ \label{f1-pert} \\
\overline{f_2}&=&\frac{\pi \eta a^2}{2}\left[-\frac{1}{\ell_1^2}
-\frac{1}{\ell_2^2}-\frac{2}{(\ell_1+\ell_2)^2}\right]
\overline{(u_1 \dot{u}_2-\dot{u}_1 u_2)}, \nonumber \\ \label{f2-pert} \\
\overline{f_3}&=&\frac{\pi \eta a^2}{2}\left[\frac{2}{\ell_1^2}
-\frac{1}{\ell_2^2}+\frac{1}{(\ell_1+\ell_2)^2}\right]
\overline{(u_1 \dot{u}_2-\dot{u}_1 u_2)}. \nonumber
\\ \label{f3-pert}
\end{eqnarray}
These forces are all proportional to the net average swimming
velocity, with proportionality constants that depend on $\ell_1$ and
$\ell_2$. Therefore, we can write a generic form
\begin{equation}
\overline{f_i}=\frac{3 \pi \eta a}{2}\;\overline{V_0}\;
g_i,\label{eq:g-def}
\end{equation}
for the forces in terms of the dimensionless factors $g_i$. These
dimensionless forces are plotted in Fig. \ref{fig:g1g2g3} as
functions of the ratio between the lengths of the two arms, which is
a measure of the asymmetry in the structure of the swimmer.

\begin{figure}
\includegraphics[width=.9\columnwidth]{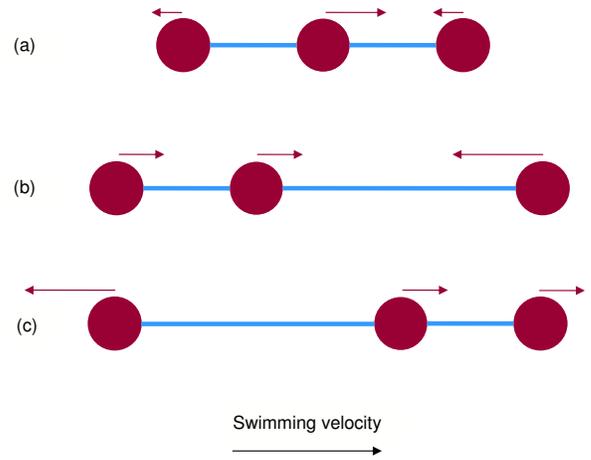}
\caption{(color online.) The distribution of the average forces
(denoted by arrows) {\em exerted on} the different spheres depending
on the relative size of the arms. Case (a) corresponds to
$\ell_1=\ell_2$, case (b) corresponds to $\ell_1 \ll \ell_2$, and
case (c) corresponds to $\ell_1 \gg \ell_2$. The middle sphere
always experiences a force that pushes it in the direction of
swimming irrespective of the structure of the swimmer. Note that the
above different cases do not correspond to the instantaneous forces
exerted on each sphere during the swimming cycle.}
\label{fig:force-distr}
\end{figure}

It is instructive to examine the limiting behaviors of the forces as
a function of the asymmetry. For $\ell_1 \ll \ell_2$ we have
$g_1=-1$, $g_2=-1$, and $g_3=2$ which means that the two closer
spheres are in line with each other and the third sphere that is
farther apart exerts the opposite force. The same trend is seen in
the opposite limit $\ell_1 \gg \ell_2$, where we have $g_1=2$,
$g_2=-1$, and $g_3=-1$. In the symmetric case where $\ell_1=\ell_2$
we have $g_1=g_3=\frac{5}{7}$ and $g_2=-\frac{10}{7}$. Note that
$g_2$ is always negative, which means that the middle sphere {\em
always} experiences a force from the fluid that pushes it in the
direction of swimming. The distribution of forces exerted on the
spheres in these three different limits is shown schematically in
Fig. \ref{fig:force-distr}.

It is interesting to note that in the symmetric case the forces are
distributed so that their net dipole moment vanishes and the first
non-vanishing moment of the forces becomes the quadrapole moment.
This will cause the net fall-off of the velocity profile at large
separations $r$ to change from $1/r^2$ (force dipoles) into $1/r^3$
(force quadrapoles). This behavior can be explained by a more
general symmetry argument \cite{gareth}.

\section{Force--Velocity Relation and Stall Force}
\label{sec:force-vel}

The effect of an external force or load on the efficiency of the
swimmer can be easily studied within the linear theory of Stokes
hydrodynamics. When the swimmer is under the effect of an applied
external force $F$, Eq. (\ref{force-free-def}) should be changed as
\begin{equation}
f_1+f_2+f_3=F.\label{force-F-def}
\end{equation}
Following through the calculations of Sec. \ref{sec:simpl} above, we
find that the following changes take place in Eqs. (\ref{v1-fi}),
(\ref{v2-fi}), (\ref{v3-fi}), and (\ref{V-def}):
\begin{eqnarray}
v_1 &\to& v_1-\frac{F}{4 \pi \eta L_1}, \label{v1-fi-F} \\
v_2 &\to& v_2-\frac{F}{6 \pi \eta a_2}, \label{v2-fi-F} \\
v_3 &\to& v_3-\frac{F}{4 \pi \eta L_2}, \label{v3-fi-F} \\
V &\to& V-\frac{1}{3}\left(\frac{1}{6 \pi \eta a_2}+\frac{1}{4 \pi
\eta L_1} +\frac{1}{4 \pi \eta L_2}\right) F. \label{V-def-F}
\end{eqnarray}
These lead to the changes
\begin{eqnarray}
{\dot L}_1 &\to& {\dot L}_1-\left(\frac{1}{6 \pi \eta
a_2}-\frac{1}{4 \pi
\eta L_1}\right) F, \label{v1-fi-F-2} \\
{\dot L}_2 &\to& {\dot L}_2-\left(\frac{1}{4 \pi \eta
L_2}-\frac{1}{6 \pi \eta
a_2}\right) F, \label{v2-fi-F-2} \\
\end{eqnarray}
in Eq. (\ref{V-full-1}), which together with correction coming from
Eq. (\ref{V-def-F}) leads to the average swimming velocity
\begin{equation}
\overline{V}(F)=\overline{V_0}+\frac{F}{18 \pi \eta
a_R},\label{V-bar-F-2}
\end{equation}
to the leading order, where $a_R$ is an effective (renormalized)
hydrodynamic radius for the three-sphere swimmer. To the zeroth
order, we have $a_R=\frac{1}{3}(a_1+a_2+a_3)$ for the general case
and there are a large number of correction terms at higher orders
that we should keep in order to be consistent in our perturbation
theory. Instead of reporting the lengthy expression for the general
case, we provide the expression for $a_1=a_2=a_3=a$, which reads
\begin{eqnarray}
\frac{1}{a_R}&=&\frac{1}{a}+\overline{\frac{1}{L_1}}+\overline{\frac{1}{L_2}}
+\overline{\frac{1}{L_1+L_2}}\nonumber \\
&-&\frac{a}{2}\overline{\left(\frac{1}{L_1}-\frac{1}{L_2}\right)^2}-\frac{a}{2}
\overline{\frac{1}{(L_1+L_2)^2}}.\label{aR}
\end{eqnarray}
We can also expand the deformation up to second order in $u_i$ and
average the resulting expression. The result of this calculation is
also lengthy and not particularly instructive, and is hence not
reported here.

The force-velocity relation given in Eq. (\ref{V-bar-F-2}), which
could have been expected based on linearity of hydrodynamics, yields
a {\em stall force}
\begin{equation}
F_s=-18 \pi \eta a_R \overline{V_0}.\label{F-s}
\end{equation}
Using the zeroth order expression for the hydrodynamic radius, one
can see that this is equal to the Stokes force exerted on the three
spheres moving with a velocity $\overline{V_0}$.

\section{Power Consumption and Efficiency}
\label{sec:power}

Because we know the instantaneous values for the velocities and the
forces, we can easily calculate the power consumption in the motion
of the spheres through the viscous fluid. The rate of power
consumption at any time is given as
\begin{equation}
{\cal P}=f_1 v_1+f_2 v_2+f_3 v_3=f_1 (-{\dot L}_1)+f_3 ({\dot
L}_2),\label{P-1}
\end{equation}
where the second expression is the result of enforcing the
force-free constrain of Eq. (\ref{force-free-def}). Using the
expressions for $f_1$ and $f_3$ as a function of ${\dot L}_1$ and
${\dot L}_2$, we find
\begin{eqnarray}
{\cal P}&=&4 \pi \eta a
\left[1+\frac{a}{L_1}-\frac{1}{2}\frac{a}{L_2}
+\frac{a}{L_1+L_2}\right]{\dot L}_1^2 \nonumber \\
&+&4 \pi \eta a \left[1-\frac{1}{2}\frac{a}{L_1}+\frac{a}{L_2}
+\frac{a}{L_1+L_2}\right]{\dot L}_2^2 \nonumber \\
&+&4 \pi \eta a
\left[1-\frac{1}{2}\frac{a}{L_1}-\frac{1}{2}\frac{a}{L_2}
+\frac{5}{2}\frac{a}{L_1+L_2}\right]{\dot L}_1{\dot L}_2,\nonumber \\
\label{P-2}
\end{eqnarray}
for $a_1=a_2=a_3=a$.

We can now define a Lighthill hydrodynamic efficiency as
\begin{equation}
\eta_{\rm L}\equiv\frac{18 \pi \eta
a_R\overline{V_0}^2}{\overline{\cal P}},\label{eta-L-def}
\end{equation}
for which we find to the leading order
\begin{eqnarray}
\eta_{\rm L}&=&\frac{9}{8}  \frac{a_R}{a} \;\frac{K^2
\;\overline{(u_1 \dot{u}_2-\dot{u}_1 u_2)}^2}{C_1\;
\overline{\dot{u}_1^2}+C_2\; \overline{\dot{u}_2^2}+C_3 \;
\overline{\dot{u}_1 \dot{u}_2}},\label{eta-L-result}
\end{eqnarray}
where $C_1=1+\frac{a}{\ell_1}-\frac{1}{2}\frac{a}{\ell_2}
+\frac{a}{\ell_1+\ell_2}$,
$C_2=1-\frac{1}{2}\frac{a}{\ell_1}+\frac{a}{\ell_2}
+\frac{a}{\ell_1+\ell_2}$, and
$C_3=1-\frac{1}{2}\frac{a}{\ell_1}-\frac{1}{2}\frac{a}{\ell_2}
+\frac{5}{2}\frac{a}{\ell_1+\ell_2}$. It is interesting to note that
for harmonic deformations (with single frequency) Eq.
(\ref{eta-L-result}) is independent of the frequency and scales like
$a^2 d^2/\ell^4$, which reflects the generally low hydrodynamic
efficiency of low Reynolds number swimmers. In this case, it is
possible to find an optimal value for the phase difference that
maximizes the efficiency \cite{feld}.

\section{Concluding Remarks}    \label{sec:conc}

We have considered the simple model of a low Reynolds number
swimmer, which is composed of three spheres that are linked by two
phantom arms with negligible hydrodynamic interaction. Assuming
arbitrary prescribed motion of the two arms, we have analyzed the
motion of the swimmer and provided explicit expressions for the
swimming velocity and other physical characteristics of the motion.

The simplicity of the model allows us to study the properties of the
swimmer in considerable details using analytical calculations. This
is a great advantage, as it can allow us to easily consider
complicated problems involving such swimmers and could hopefully
lead to new insights in the field of low Reynolds number locomotion.
An example of such studies has already been performed by Pooley {\em
et al.} who considered the hydrodynamic interaction of two such
swimmers and found a rich variety of behaviors as a function of
relative positioning of the two swimmers and their phase coherence
\cite{gareth}. This can be further generalized into a multi-swimmer
system, and the collective floc behavior of such systems can then be
studied using a ``realistic'' model for self-propellers at low
Reynolds number that is faithful to the rules of the game. Knowing
something about the internal structure of a swimmer will also allow
us to study the synchronization problem more systematically
\cite{sync}.

\acknowledgements

We would like to thank T.B. Liverpool, A. Najafi, and J. Yeomans for
discussions.

\appendix

\section{Swimming Velocity for Arbitrary Deformations}   \label{app:V-full}

For the general case of a three-sphere swimmer based on the
schematics in Fig. \ref{fig:3SS-schem}, we obtain the average
swimming velocity to the leading order as
\begin{widetext}
\begin{eqnarray}
V_0&=&\frac{(a_1-a_2) (a_2+a_3)}{3 a_2 (a_1+a_2+a_3)}\left[1
+\frac{3}{2} \left(\frac{a_1 a_2}{a_2-a_1}\right)
\left(\frac{1}{L_1+L_2}-\frac{1}{L_2}\right) -3 \left(\frac{a_2
a_3}{a_2+a_3}\right) \frac{1}{L_2} \right.\nonumber \\
&& \hskip3cm \left.+\frac{3}{a_1+a_2+a_3}
\left(\frac{a_2 a_3}{L_2}+\frac{a_1 a_2}{L_1}+\frac{a_3 a_1}{L_1+L_2}\right) \right] {\dot L}_1  \nonumber \\
&+&\frac{a_3 (a_1-a_2)}{3 a_2 (a_1+a_2+a_3)}\left[1 +\frac{3}{2}
\left(\frac{a_1 a_2}{a_2-a_1}\right)
\left(\frac{1}{L_1+L_2}-\frac{1}{L_2}\right) -\frac{3}{2}
\left(\frac{a_2}{L_1}+\frac{a_2}{L_2}-\frac{a_2}{L_1+L_2}\right)\right.\nonumber \\
&& \hskip3cm \left.+\frac{3}{a_1+a_2+a_3}
\left(\frac{a_2 a_3}{L_2}+\frac{a_1 a_2}{L_1}+\frac{a_3 a_1}{L_1+L_2}\right) \right] {\dot L}_2  \nonumber \\
&+&\frac{a_1 (a_2-a_3)}{3 a_2 (a_1+a_2+a_3)}\left[1 +\frac{3}{2}
\left(\frac{a_2 a_3}{a_2-a_3}\right)
\left(\frac{1}{L_1+L_2}-\frac{1}{L_1}\right)-\frac{3}{2}
\left(\frac{a_2}{L_1}+\frac{a_2}{L_2}-\frac{a_2}{L_1+L_2}\right) \right.\nonumber \\
&& \hskip3cm \left.+\frac{3}{a_1+a_2+a_3}
\left(\frac{a_2 a_3}{L_2}+\frac{a_1 a_2}{L_1}+\frac{a_3 a_1}{L_1+L_2}\right) \right] {\dot L}_1  \nonumber \\
&+&\frac{(a_2-a_3) (a_1+a_2)}{3 a_2 (a_1+a_2+a_3)}\left[1
+\frac{3}{2} \left(\frac{a_2 a_3}{a_2-a_3}\right)
\left(\frac{1}{L_1+L_2}-\frac{1}{L_1}\right) -3 \left(\frac{a_1
a_2}{a_1+a_2}\right) \frac{1}{L_1} \right.\nonumber \\
&& \hskip3cm \left.+\frac{3}{a_1+a_2+a_3} \left(\frac{a_2
a_3}{L_2}+\frac{a_1 a_2}{L_1}+\frac{a_3 a_1}{L_1+L_2}\right) \right]
{\dot L}_2.  \label{V-full-1}
\end{eqnarray}
\end{widetext}
This expression can be used in numerical studies of the swimming
motion for arbitrarily large deformations and geometric
characteristics.


\begin{thebibliography}{99}
\bibitem{note}
We would like to point out that in Ref. \cite{3SS} a relation is
proposed for the average swimming velocity which numerically fits
best to a large portion of the parameters encompassing small and
large deformations, rather than attempting to systematically perform
a perturbative calculation. This is also discussed in Ref.
\cite{earl}.
\end{thebibliography}
\end{document}